# 3D-Imaging and Quantification of Magnetic Nanoparticle Uptake by Living Cells

Hendrik Paysen[a*], Norbert Loewa[a], Anke Stach[b], James Wells[a], Olaf Kosch[a], Shailey Twamley[b], Marcus R. Makowski[c], Tobias Schaeffter[a], Antje Ludwig[b,c,d] and Frank Wiekhorst[a]

[a] *Physikalisch-Technische Bundesanstalt, Berlin, Germany*

[b] *Charité – Universitätsmedizin Berlin, Medizinische Klinik mit Schwerpunkt Kardiologie und Angiologie; Campus Mitte*

[c] *Charité – Universitätsmedizin Berlin, Klinik für Radiologie*

[d] *DZHK (German Centre for Cardiovascular Research), partner site Berlin, Berlin, Germany*

[*] *Corresponding author, email: hendrik.paysen@ptb.de*

**Abstract** – **Magnetic particle imaging (MPI) is a non-invasive, non-ionizing imaging technique for the visualization and quantification of magnetic nanoparticles (MNPs). The technique is especially suitable for cell imaging as it offers zero background contribution from the surrounding tissue, high sensitivity, and good spatial and temporal resolutions. Previous studies have demonstrated that the dynamic magnetic behaviour of MNPs changes during cellular binding and internalization. In this study, we demonstrate how this information is encoded in the MPI imaging signal. Through MPI imaging we are able to discriminate between free and cell-bound MNPs in reconstructed images. This technique was used to image and quantify the changes that occur in-vitro when free MNPs come into contact with cells and undergo cellular-uptake over time. The quantitative MPI results were verified by a phenanthroline assay. The results showed a mean relative difference of 23.8% for the quantification of cell-bound MNPs. The insights gained from such observations provide a new window into fundamental biological processes and associated pathological changes occurring at a cellular level. This technique could therefore offer new opportunities for the early diagnosis of inflammatory diseases.**

*Graphical Abstract:*

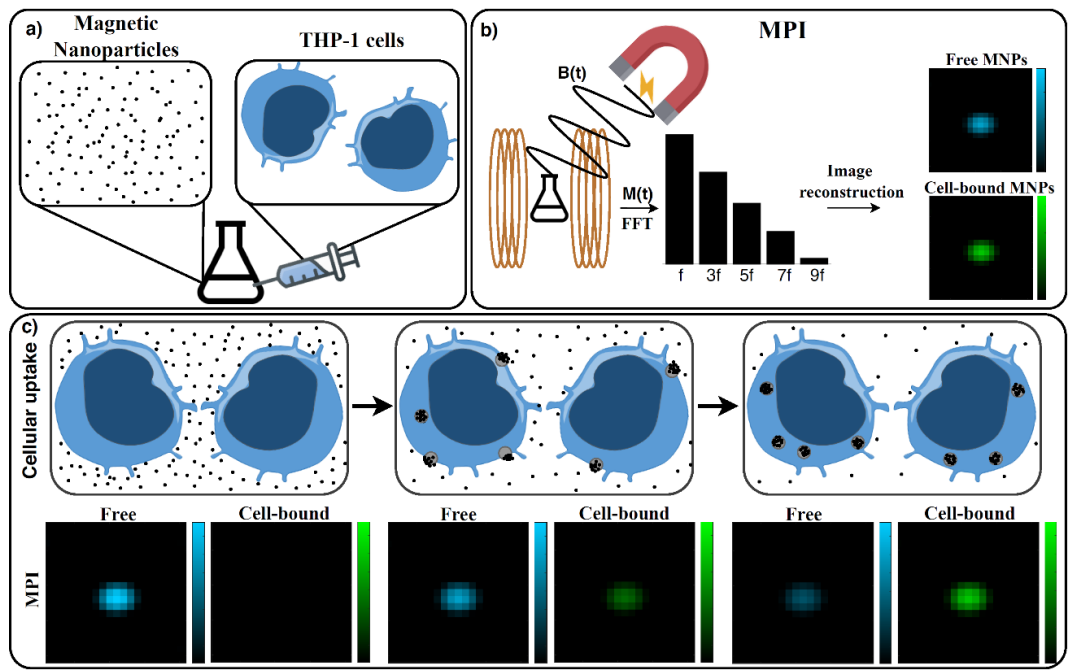

# INTRODUCTION

Magnetic particle imaging (MPI) is a non-invasive technique capable of determining the spatial distribution of magnetic nanoparticles (MNPs) both in-vivo and in-vitro [1]. MPI achieves imaging and quantification detecting the non-linear dynamic magnetic response of MNPs exposed to multiple superimposed static and dynamic magnetic fields. No background signals are generated by bone or tissue. The technique uses non-ionizing radiation and non-toxic nanoparticles to prevent tissue damage. MPI shows great potential for different biomedical applications, such as angiography, stem cell tracking, diagnosis of inflammatory diseases and cancer [2]–[11]. In inflammation-associated diseases, including cancer, MNPs accumulate preferentially in diseased tissue as a result of leaky vasculature thereby enabling imaging with MRI and MPI [12]–[16]. In diseased tissue, MNPs accumulate mainly in macrophages [17], [18]. These phagocytic cells are a hallmark of tissue inflammation and their quantity is considered a marker of the severity of the disease [19]–[21].

The magnetic response generated by MNPs is severely influenced by their local environment. In this respect, the changing magnetic properties of MNPs induced by their interaction with cells are a relevant factor for MPI. Previous studies using magnetic particle spectroscopy (MPS), have described and quantified the changes that occur in the dynamic magnetization of certain MNP systems upon interaction with living cells [22]–[26]. These effects may be caused by a variety of factors including the aggregation of particles in the surrounding solution, within the extra-cellular matrix or within various intracellular compartments. "Size-filtering" during cellular uptake and an increase in dipole-dipole interactions caused by a lower separation and mobility of the MNPs may also be influencing this phenomenon [25]–[28]. Typically, these signal changes lead to a deterioration of MPI images, resulting in lower sensitivity, decreased spatial resolution and more imaging artifacts [26], [28]–[30]. However, if the unique magnetic signal patterns induced by the interaction of MNPs with cells are known, they can be incorporated into the image reconstruction to negate or minimize these negative effects. Additionally, this enables the possibility to search and separate these unique signal patterns generated by MNPs in specific environmental conditions. Such a technique, known as multi-color or multi-contrast MPI, has been previously demonstrated by separating the signals of different MNP systems under varying parameters such as temperature or viscosity [30]–[32].

The aim of this work was to test the potential of imaging the internalization of MNPs into living cells with MPI. It was hypothesized, that based on different magnetic signals, the particle distributions of free vs cell-bound MNPs can be separated using MPI image reconstruction. Furthermore, due to the high temporal resolution of MPI, the processes involved in the cellular uptake of MNPs were imaged and quantified.

# MATERIALS AND METHODS

*MPI acquisition*

All MPI measurements were performed on a commercial, preclinical MPI scanner (Bruker MPI 25/50 FF) based on the field-free point (FFP) approach. The FFP is generated by a static magnetic gradient field with gradient strengths 0.6/0.6/1.2 T/m (*x*/*y*/*z*-direction). By superimposing oscillating magnetic drive fields with field amplitudes of 12/12/12 mT and frequencies 24.5/26.0/25.3 kHz the FFP is moved on a Lissajous trajectory through the field of view (FOV) of size 40/40/20 mm. Based on these magnetic field settings and the nonlinear dynamic magnetic properties of MNPs, higher harmonics of the excitation frequencies are generated and measured by a gradiometric receive coil [33]. The spectral patterns of the MPI signal (*u*) depend on the MNP system, the environmental conditions and the spatial distribution of the MNPs within the FOV. A 3D image reconstruction of the MNP distribution ($c$) is performed by solving the following least-squares problem using the Kaczmarz algorithm with Tikhonov regularization (regularization parameter $\lambda$):

$$\|\boldsymbol{Sc} - \boldsymbol{u}\|^2 + \lambda \|\boldsymbol{c}\|^2 \rightarrow min.$$

The SF ($\boldsymbol{S}$) contains information on how the environmental conditions and spatial position affect the measured MPI signal $\boldsymbol{u}$. The SFs used in this study were acquired experimentally by measuring small reference samples of MNPs with fixed environmental conditions at multiple positions all over the scanner's FOV. For this approach to be accurate, it is necessary that all environmental conditions influencing the MPI signal during a measurement are the same as the conditions during the SF acquisition. If the environmental conditions of the MNPs change, it must be considered in the image reconstruction process. This can be achieved by acquiring multiple SFs ($\boldsymbol{S_1}$ and $\boldsymbol{S_2}$) of MNPs at different environmental conditions. Thus, a combined SF can be used to reconstruct multiple particle distributions ($\boldsymbol{c_1}$ and $\boldsymbol{c_2}$) of MNPs at given environmental conditions by adapting the least squares problem:

$$\left\| [\boldsymbol{S_1}\boldsymbol{S_2}] \begin{bmatrix}\boldsymbol{c_1}\\\boldsymbol{c_2}\end{bmatrix} - \mathbf{u} \right\|^2 + \lambda \left\| \begin{bmatrix}\boldsymbol{c_1}\\\boldsymbol{c_2}\end{bmatrix} \right\| \rightarrow min.$$

Assuming that changes in the dynamic magnetization induced by interactions of cells and MNPs can be described with a simple two-state system, two SFs were acquired under same field settings with a fluid (8 µL Synomag, iron oxide MNP, product nr. 103-02-301, LOT: 05218103-03 iron concentration c(Fe)=2.75 mg/ml, micromod Nanopartikeltechnologie GmbH Rostock) diluted in phosphate-buffered saline (PBS)) and cell sample ($2 \cdot 10^6$ THP-1 cells incubated with Synomag), respectively. This enables the reconstruction of two particle distributions representing the free and cell-bound MNP distributions. The choice of reconstruction parameters heavily influences the quantitative and qualitative MPI images. Therefore, we varied the reconstruction parameters over a certain range, taking a priori knowledge of the particle distributions into account to determine an optimal parameter set. A detailed description of this procedure, and a list of all measurement and reconstruction parameters used in this study can be found in the supplement.

*Cell cultivation and sample preparation*

THP-1 cells (human acute monocytic leukemia cell line) obtained from the ATCC (Wesel, Germany) were cultured in suspension in a humidified incubator at 37 °C with a 5% $CO_2$ concentration in RPMI medium 1640 (Invitrogen, Karlsruhe, Germany). Culture medium was supplemented with 10% fetal calf serum (FCS, Biochrom, Berlin, Germany), 100 U/ml penicillin, 100 µg/ml streptomycin (Invitrogen), and 2 mM L-glutamine (Invitrogen). Exact cell numbers were determined with a hemocytometer.

The cell sample for system function acquisition was prepared as follows: $2 \cdot 10^6$ THP-1 cells suspended in 2 mL of RPMI containing 1% FCS were treated for 3 hours with 0.5 mM Synomag. After incubation cells were centrifuged for 3 min at 200 g and pellets were washed three times with 1 mL PBS to remove all unbound MNPs. After washing, resuspended cells were filled into a cubic 2 mm³ container (Bruker Biospin, Germany). Centrifugation for 3 min at 200 g led to the sedimentation of the cells in the lower chamber of the container. After SF acquisition the iron content of the pellet was quantified using the 1,10-phenanthroline-based iron assay as previously described [15].

For the colorimetric quantification of Synomag uptake similar conditions as for the quantification with MPI were used. As described in the experimental setup: The indicated numbers of THP-1 cells were suspended in 100 µL PBS, and added in a 1.5 mL Eppendorf tube to 40 µL of Synomag diluted in PBS with an iron concentration of 2.75 mg/mL. After incubation cells were centrifuged for 3 min at 200 g. Supernatants were collected and evaporated to dryness using a Speedvac (Savant Instruments, NY, USA). Pellets were washed twice with 1 mL PBS and centrifuged for 3 min at 200 g. Iron content in the sediment of the supernatants and of the pellets was quantified using the 1,10-phenanthroline-based iron assay.

*Experimental setup: in-vitro imaging*

A vessel containing 40 µL of Synomag diluted in PBS was prepared with an iron concentration of 2.75 mg/mL and mounted to a sample holder, which can be moved to the centre of the MPI FOV. A set of THP-1 cells with varying number of cells in the range $0 - 10^6$ were suspended in 100 µL PBS, drawn up into a syringe and connected with a tube to the vessel. In each MPI data acquisition, the scanner ran continuously for 29 minutes, obtaining a new averaged full 3D dataset every 2.15 s. In each acquisition, the initial measurements were of the empty scanner only; these data were used for background correction. After two minutes the vessel was moved to the FOV centre, measuring only MNPs diluted in PBS without contact to cells. After one more minute the THP-1 cells were injected to the MNPs (time point defined as $t = 0$). For each time point the free ($c_1$) and cell-bound ($c_2$) MNP amounts were reconstructed. The reconstructed MPI signals were summed in the region of interest (ROI) around the nominal sample positions and converted into iron masses of the free ($m_{\text{Fe},free}$) and cell-bound ($m_{\text{Fe},cell}$) MNP distribution. The relative deviation of the total iron amount compared to the initially determined iron mass determined before cell-injection ($m_{\text{Fe},0}$) was calculated $\Delta m_{\text{Fe}} = \frac{(m_{\text{Fe},free} + m_{\text{Fe},cell}) - m_{\text{Fe},0}}{m_{\text{Fe},0}}$ for each frame.

**RESULTS**

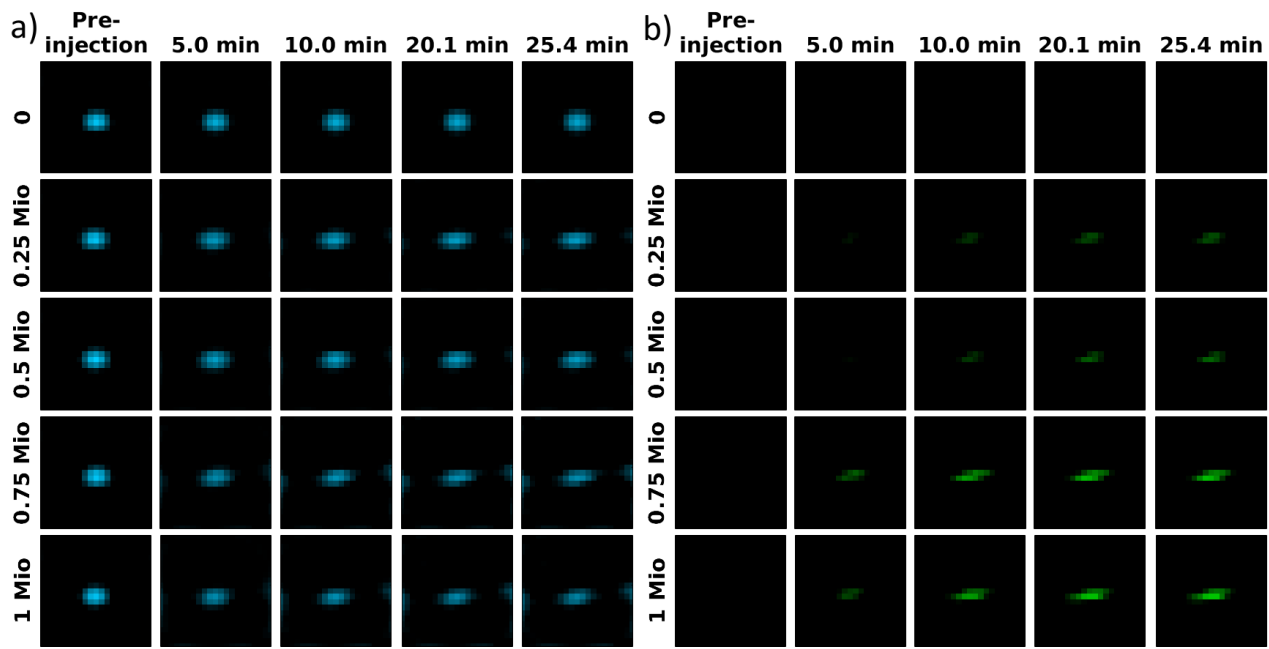

*Figure 1: Reconstructions as a function of time of free (a) and cell-bound (b) MNPs in contact with THP-1 cells. The first column displays the reconstruction before injection of the cells (number of cells displayed on the left), in which only MNPs diluted in PBS were imaged. The following columns show representative time frames after the injection. Displayed are summed up intensities along the z-axis to visualize the 3D datasets. The same scaling was used for all images.*

Figure 1 shows MPI reconstructions of free (a) and cell-bound (b) MNP-distributions of representative time frames before and after injection of THP-1 cells. The sum of the intensities along the z-axis is presented in each case, to visualize the 3D dataset in a 2D image. The control measurement performed by injecting PBS without any THP-1 cells, showed no significant influence on the reconstructed images of free MNPs and no detection of cell-bound MNPs. Qualitatively, there were no significant differences detected in the reconstructions of free MNPs at different time steps. These were only visible when high numbers of cells were injected. Minor boundary artifacts can be observed starting at about 10 min after injection. Increasing the number of injected cells, led to a formation of cell-bound MNPs at the same position as the free MNPs with increasing intensities over time.

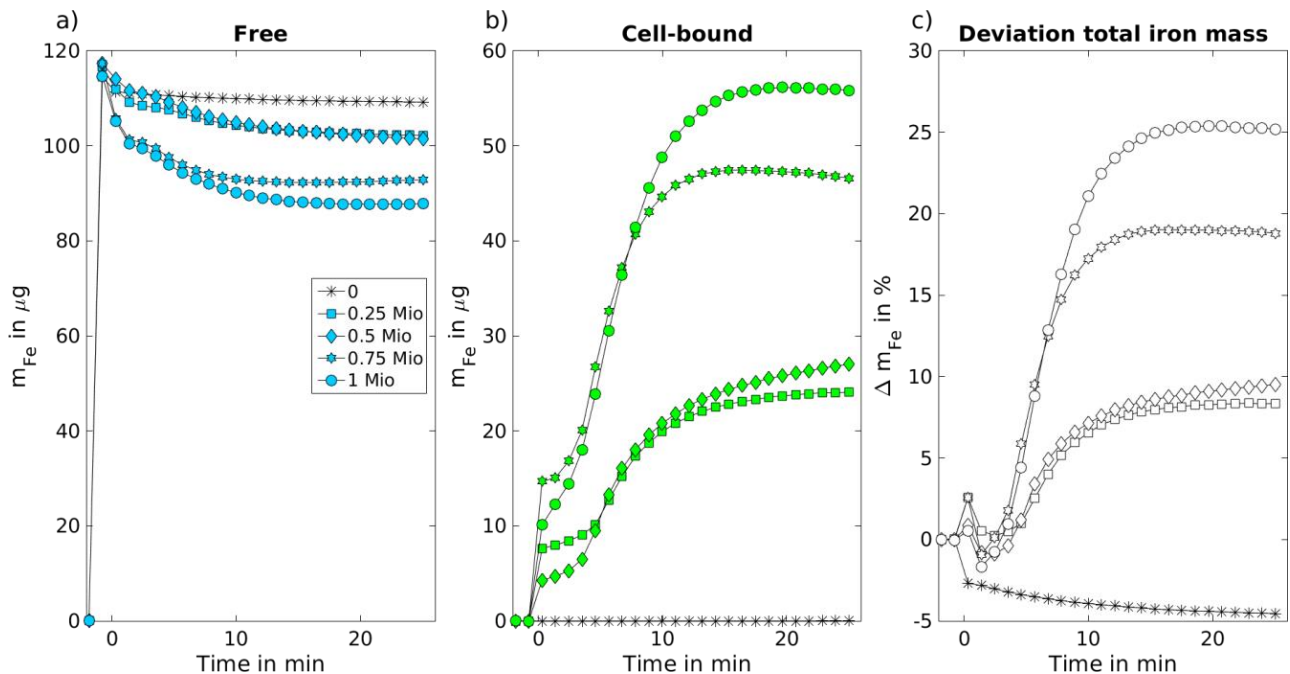

*Figure 2: Quantified iron quantities of free (a) and cell-bound MNPs during initial MNP-THP-1 cell contact. With an increasing number of injected cells, a stronger increase of cell-bound MNPs and decrease of free MNPs was detected. c) displays the relative deviation of the total iron amount (free+cell-bound MNPs) from the initially quantified amount of free MNPs before the injection. For better visualization, only a fraction of all available data points was incorporated in the figure.*

The quantified iron amounts of free and cell-bound MNPs were extracted from the reconstructions and are displayed over time in Figure 2a) and 2b). After cell-injection, the amount of free MNPs decreases over time. The more cells were injected, the stronger the measured decrease of free MNPs. Simultaneously, the quantified values of cell-bound MNPs increases over time, also scaling with the number of injected cells. Saturation was reached after 10-15 minutes for all measurements. For each measurement, the quantified values of cell-bound MNPs were determined $\approx$ 30 minutes after injection and the average loading per cell was calculated to be $m_{Fe} = (67 \pm 19)$ pg.

The relative deviation of the total iron amount ($\Delta m_{Fe}$) is displayed over time in Figure 2c). Overall, the deviation increases with the number of injected cells. A maximum deviation was determined after injecting $10^6$ THP-1 cells, overestimating the total iron mass by 25.4%. The control experiment, performed by injection of PBS without cells, showed no significant changes in the quantified free and cell-bound iron masses. However, there was a small decrease below 5% of the free iron mass within the whole acquisition time. All reconstructed MPI images can be found as animated versions together with the quantification results in the supplementary data.

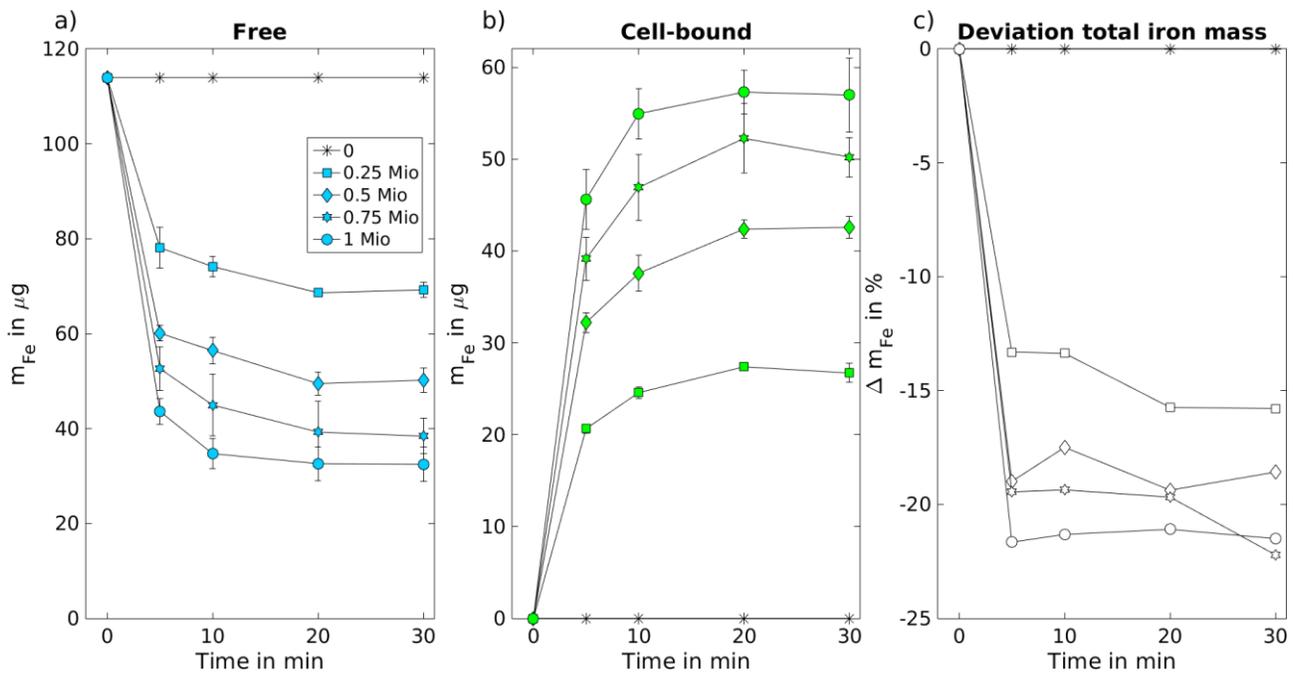

*Figure 3: Quantified iron amounts acquired via the phenanthroline-based iron assay method for varying incubation times. Each data point represents the mean value from three independent measurements with the standard deviation visualized as error bars. Note that the lines connecting the dots do not represent measurement data and are only shown for better visualization. Due to the required washing steps during the phenanthroline-based iron assay method, some MNPs are lost, leading to an underestimation of the total iron mass by up to -22%, especially regarding the free MNPs.*

The quantified iron amounts of the reference experiments, performed using the phenanthroline-based iron assay method, are displayed in Figure 3. Each data point represents the averaged iron mass acquired from three independent measurements with the standard deviation visualized as error bars. Qualitatively, similar behaviour as in the MPI measurements can be observed. The free amounts of MNPs were decreasing over time until a saturation is reached after about 10-15 minutes. Simultaneously, the amount of cell-bound MNPs increased. The higher the number of injected cells, the stronger this effect was observed. After 30 minutes of MNP-cell exposure the average iron mass per cell was determined to be $m_{Fe} = (80 \pm 22)$ pg per cell. Figure 3c displays the relative deviation of the total iron mass, showing an underestimation of the total iron amount of up to -22%.

## DISCUSSION

The central finding of this study is, that the distribution of free and cell-bound Synomag can be separated by our extended MPI image reconstruction based on different magnetic signals. This was achieved by incorporating the different magnetic signals generated by Synomag diluted in water and after cellular internalization into the MPI image reconstruction. This allowed for both the visualization and quantification of the cellular uptake of Synomag by THP-1 cells with a high temporal resolution of 2.15 s.

A decrease in the amount of reconstructed free MNPs and an increase in the amount of cell-bound MNPs over time supports the assumption that MNPs were accumulating in the cells. A similar pattern was observed when MPI results were qualitative compared with the results acquired with the reference experiment (phenanthroline-based iron assay). Both show a consistent decrease of free and increase of cell-bound MNPs until saturation is reached after about 10-15 minutes. This rapid cellular uptake of electrostatically stabilized MNPs is in accordance with previous studies using different particle systems [23]. There is a relative difference (23.8%) between the quantified values of cell-bound MNPs determined by MPI and the phenanthroline-based iron assay. Stronger differences were observed when comparing the quantified values of free MNPs. These were mainly the result of the loss of MNPs during the required washing and vacuum evaporation steps associated with the phenanthroline-based iron assay. Hence, the amount of free MNPs was underestimated.

The deviation of the total iron amount determined by MPI increased over time and correlated with the number of injected cells. These deviations can only partly be attributed to technical issues, such as heating of hardware components and thermal drifts of the background signal, which caused small deviations below 5% in the control experiment (as determined from injection of PBS without cells). The impact of thermal drifts could be minimized by an advanced background signal removal technique [34], [35]. We assume the main reason for the deviation was caused by MNPs whose magnetic properties deviate from the states "free MNP" and "cell-bound MNPs", which we used in this proof-of-principle study. The detection of signals generated by additional states leads to quantification errors and also explains the imaging artifacts occurring at the edges of the FOV after 10 minutes in some of the reconstructed images [30], [36]. We have previously shown, that MNPs undergo multiple changes of magnetic behaviour during cellular uptake [23], [24]. At initial contact, MNPs are interacting with the pericellular matrix, a highly hydrated mesh consisting mainly of complex carbohydrates, such as glycosaminoglycans. This is followed by the initiation of a process known as endocytosis. Endocytosis is the process of internalizing extracellular substances which are then processed by a variety of intracellular vesicles such as endosomes and lysosomes. Later, MNPs are increasingly packed densely inside endosomes and finally degraded in the acidic environment of the endolysosomes [37]. All these stages seemed to be characterized by a distinct magnetic signature that can theoretically each be registered by a separate SF acquisition [23], [24]. The creation of such a SF matrix would complicate the image reconstruction from a mathematical point of view but could be adapted to the respective experimental research question.

A different approach to include multiple states in the MPI image reconstruction would be to take the reconstructed signals gained from a two-state system as presented here. The respective MPI signals could then be correlated with a priori knowledge of the measured sample to generate a calibration curve of the cellular uptake process. A similar approach was presented for viscosity quantification using multi-contrast MPI [32].

It was shown, that MPI can not only be used to image and quantify the spatial distribution of MNPs, but also to extract functional information about the environmental conditions such as the binding state based on the simple assumption of a two-state model. Overall imaging the cellular uptake process was possible for cell numbers down to $2.5 \cdot 10^5$ with a temporal resolution of 2.15 s. This technique can be adapted for other cell types and MNP systems. This would permit for nanoscale imaging of MNP cellular internalization without damaging or destroying the cells. Further improvements can be achieved by adapting the reconstruction process and by extending the two-state model as discussed above.

## Conclusion

This study demonstrates the potential of MPI to separate the signals of free and cell-bound MNPs, within a mixed sample. This technique can be used to image and quantify the uptake of MNPs into cells starting directly from initial MNP-cell contact thereby providing information about the dynamics of cellular uptake. It is assumed that the uptake dynamics of MNPs *in vivo* correlates with pathological changes of diseased tissue, in particular with the permeability of the vasculature and the infiltration of immune cells. Therefore, further development of this technique is anticipated to result in highly interesting opportunities for future diagnostic applications.

**Acknowledgment**

*This project was supported by the DFG research grants "AMPI: Magnetic particle imaging: Development and evaluation of novel methodology for the assessment of the aorta in vivo in a small animal model of aortic aneurysms" (grant SHA 1506/2-1), "quantMPI: Establishment of quantitative Magnetic Particle Imaging (MPI) application oriented phantoms for preclinical investigations" (grant TR 408/9-1), "BIOQIC" (GRK2260) and "Matrix in Vision" (SFB 1340/1 2018, no 372486779, projects A02 and B02). We would like to acknowledge the support of Cordula Grüttner from Micromod with their expertise in fabrication of magnetic nanoparticles.*

**Author contributions**

H.P., N.L., A.L. and F.W. contributed to critical discussion regarding the experimental design.  H.P., A.S. and O.K. contributed to the MPI measurements. A.S., S.T. and A.L. contributed to preparation of cell samples. H.P., N.L., J.W., S.T., M.M., T.S., A.L. and F.W. contributed to critical discussion regarding the measurement results. H.P. wrote the manuscript.

**Additional Information**

Competing interest: The authors have declared that no competing interest exists.

Abbreviations: FCS: fetal calf serum; FFP: field-free point; FOV: field of view; MNP: magnetic nanoparticles; MPI: magnetic particle imaging; MPS: magnetic particle spectroscopy: MRI: magnetic resonance imaging; PBS: phosphate-buffered saline; ROI: region of interest; SF: system function;

**SUPPLEMENTARY METHODS**

*MPI measurement parameters*

| System function parameters ||
|---|---|
| **Parameter** | **Value/description** |
| **Drive fields** | 12/12/12 mT |
| **Gradient strength** | 0.6/0.6/1.2 T/m |
| **System function grid** | 21x21x12 |
| **FOV** | 42x42x24 mm³ |
| **Averages** | 100 |
| **Background correction** | Background increment 21; 5 BG repetitions |
| **Data acquisition software** | Paravision 6.0 (Bruker) |
| **Data processing** | Matlab 2015a (Mathworks) |

| MPI measurement parameters in-vitro study ||
|---|---|
| **Parameter** | **Value/description** |
| **Drive fields** | 12/12/12 mT |
| **Gradient strength** | 0.6/0.6/1.2 T/m |
| **Averages** | 100 (2.15 s acquisition time) |
| **Repetitions** | 800 (total acquisition time 29 min) |
| **Receive unit** | Gradiometric coil |
| **Background correction** | Subtraction of empty measurements |
| **Data acquisition software** | Paravision 6.0 (Bruker) |
| **Data processing** | Matlab 2015a (Mathworks) |

*Reconstruction parameters*

A decisive role for MPI image reconstruction is the choice of the reconstruction parameters. In general, the MPI imaging equation is formulated as an underdetermined linear system. Therefore, the existence of a unique solution cannot be guaranteed. Hence, a least-squares problem is formulated to guarantee a solution, which is solved by the Kaczmarz algorithm with Tikhonov regularization. However, the number of frequency components used in the reconstruction, the number of iterations and the regularization parameter are variable and might have severe influence on the qualitative and quantitative values of the reconstructed images. In most MPI studies these parameters are chosen manually. Here, we performed a reconstruction parameter study in combination with a priori knowledge of the particle distribution to determine a full set of reconstruction parameters. The final reconstruction parameters are then used for the reconstructions of all measurement data to allow a quantitative comparison. In the following the procedure will be explained.

A time frame before injection of the cells was chosen for which the quantitative particle distribution is known. After background correction and truncation of the frequency components (>60 kHz), multiple image reconstructions were performed using the same measurement data with varying reconstruction parameters. The number of frequency components (FCs) was varied between $25 - 2000$, sorted by the SNR of the SF, the relative regularization parameter $\lambda$ (see [1]) was varied between $5 \cdot 10^{-5} - 5 \cdot 10^0$ and the number of iterations between $1 - 1000$. The intensities in a region of interest (ROI) around the nominal sample position were summed up and converted into the respective iron masses. Since the particle distribution is known, the difference of the reconstructed iron mass and the nominal iron mass can be calculated $\Delta m_{Fe} = m_{Fe,reco} - m_{Fe,nominal}$ for the free and cell-bound particle distributions respectively, which is visualized as a function of the reconstruction parameters in figure S1. The reconstruction parameters will be chosen to minimize $\Delta m_{Fe}$ for both particle distributions. The final parameters are listed in the table below.

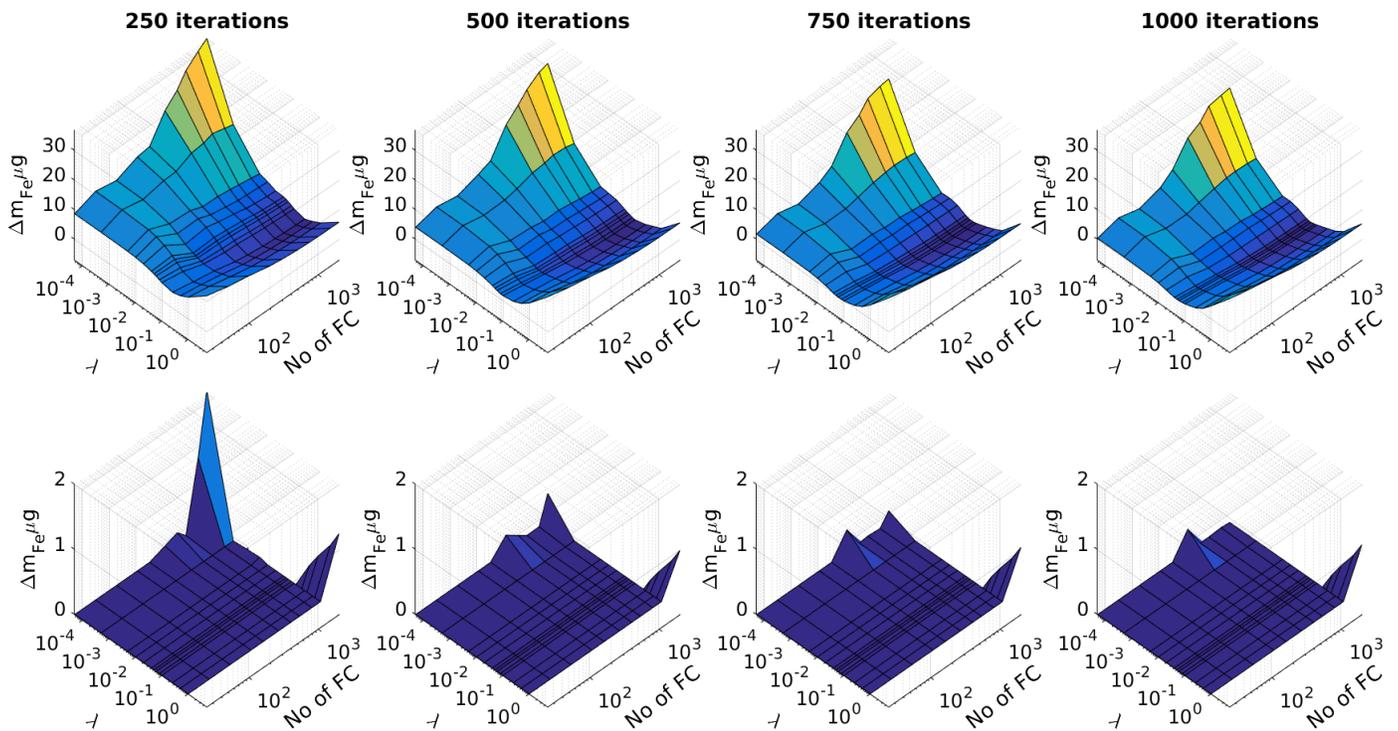

*Figure S1: Visualization of reconstructed iron masses of free and cell-bound particle distributions determined in a ROI for the in-vitro study. Based on a priori knowledge of the nominal particle distributions, the difference to the reconstructed iron mass $\Delta m_{Fe}$ was calculated and is minimized to determine the final reconstruction parameters used for all measurements.*

| MPI reconstruction parameters in-vitro study ||
|---|---|
| **Parameter** | **Value/description** |
| **Number of FC** | 500 |
| **Reconstruction algorithm** | Kaczmarz |
| **Regularization** | 0.04 |
| **Iterations** | 1000 |

**Supplementary references**